\documentclass[10pt]{article}
\usepackage{a4wide}

\newcommand{\dd}{{\rm{d}}} 
\newcommand{\rovno}{\!\!\!\!& = &\!\!\!\!}
\newcommand{\erovno}{\!\!\!\!& = &\!\!\!\!}

\begin{document}

\title{Absence of gyratons in the Robinson--Trautman class}

\author{
R.~\v{S}varc\thanks{{\tt robert.svarc@mff.cuni.cz}} \ and \ J.~Podolsk\'y\thanks{{\tt podolsky@mbox.troja.mff.cuni.cz}}\\ \\
Institute of Theoretical Physics,
Charles University in Prague,\\ Faculty of Mathematics and Physics, \\
V~Hole\v{s}ovi\v{c}k\'ach 2, 18000 Prague 8, Czech Republic.\\ \\
}

\maketitle

\begin{abstract}
We present the Riemann and Ricci tensors for a fully general non-twisting and shear-free geometry in arbitrary dimension $D$. This includes both the non-expanding Kundt and expanding Robinson--Trautman family of spacetimes. As an interesting application of these explicit expressions we then integrate the Einstein equations and prove a surprising fact that in any $D$ the Robinson--Trautman class does not admit solutions representing gyratonic sources, i.e., matter field in the form of a null fluid (or particles propagating with the speed of light) with an additional internal spin. Contrary to the closely related Kundt class and pp-waves, the corresponding off-diagonal metric components thus do not encode the angular momentum of some gyraton. Instead, we demonstrate that in standard ${D=4}$ general relativity they directly determine two independent amplitudes of the Robinson--Trautman exact gravitational waves.
\end{abstract}

\vfil\noindent
PACS class:  04.20.Jb, 04.30.--w, 04.50.--h, 04.40.Nr


\vfil
\eject

\section{Introduction}
The Robinson--Trautman class of spacetimes, discovered more than fifty years ago \cite{RobTra60,RobTra62}, is one of the most fundamental families of exact solutions to Einstein's field equations. Geometrically, it is defined by admitting a geodesic, shear-free, twist-free but \emph{expanding} null congruence. This group of spacetimes contains many important vacuum solutions, in particular Schwarzschild-like static black holes, accelerating black holes ($C$-metric) and radiative spacetimes of various algebraic types. It also admits a cosmological constant, electromagnetic field or pure radiation, as in the case of the Vaidya metric or Kinnersley photon rockets. More details and a substantial list of references can be found in chapter~28 of \cite{Stephani:2003} or chapter~19 of \cite{GriffithsPodolsky:2009}.

In \cite{PodOrt06} the Robinson--Trautman family of solutions was extended to higher dimensions~$D$ in the case of empty space (with any value of the cosmological constant) and for aligned pure radiation. Interestingly, there are great differences with respect to the usual ${D=4}$ case (see also \cite{Ortaggio07}). Aligned electromagnetic fields were subsequently also incorporated into the Robinson--Trautman higher-dimensional spacetimes within the Einstein--Maxwell theory \cite{OrtPodZof08}, and an additional Chern--Simons term for ${D\ge 5}$ odd dimensions was also considered. The results were recently summarized in the review work \cite{OrtaggioPravdaPravdova:2013} on algebraic properties of spacetimes of higher dimensions.

The complementary \emph{non-expanding} Kundt class of twist-free and shear-free geometries also admits explicit vacuum solutions with an arbitrary cosmological constant, electromagnetic fields and pure radiation (null fluid), see chapter~31 of \cite{Stephani:2003} or chapter~18 of \cite{GriffithsPodolsky:2009} for summary concerning the Einstein theory in ${D=4}$. The corresponding extensions to higher dimensions were presented in the work \cite{PodolskyZofka:2009}. Interestingly, the whole Kundt class also admits spacetimes representing null fields of \emph{gyratonic matter} with internal spin/helicity. It turns out that the angular momentum of such rotating sources is encoded in the non-diagonal metric functions.

This observation was made by Bonnor already in 1970 \cite{Bonnor:1970b,Griffiths:1972}. He studied both the interior and the exterior field of a ``spinning null fluid'' in the class of axially symmetric \emph{pp}-wave spacetimes which are the simplest representatives of the Kundt family.
In the natural coordinates of non-twisting geometries (see section~\ref{sec_geom}), the energy-momentum tensor in the interior region is phenomenologically described by the radiation energy density $T_{uu}$ and by the components $T_{up}$ representing the spinning character of the source (its non-zero angular momentum density). Spacetimes with such localized spinning sources moving at the speed of light were independently rediscovered and investigated (in four and higher dimensions) in 2005 by Frolov and his collaborators who called them gyratons \cite{FrolovFursaev:2005,FrolovIsraelZelnikov:2005}. These \emph{pp}-wave-type gyratons were later studied in greater detail and generalized to include a negative cosmological constant \cite{FrolovZelnikov:2005}, electromagnetic field \cite{FrolovZelnikov:2006},
and various other settings including non-flat backgrounds. An extensive summary can be found in \cite{KrtousPodolskyZelnikovKadlecova:2012}. This recent work presents and investigates gyratons in a fully general class of Kundt spacetimes in any dimension.

In fact, all the so far known spacetimes with gyratonic sources belong to the Kundt class. The question thus arises: Is it possible to find gyratons in other geometries? The most natural candidate is clearly the Robinson--Trautman family because it shares the non-twisting and shear-free property and in ${D=4}$ it admits a similar algebraic structure. It differs only in having a non-vanishing expansion of the geometrically privileged null congruence.

This is the purpose of the present paper: We systematically study the possible existence of Robinson--Trautman gyratonic solutions (in any dimension) which would be analogous to those known in the Kundt class. First, in section~\ref{sec_geom} we present the general form of the non-twisting shear-free line element and all its components of the Christoffel symbols, the Riemann and the Ricci tensors. In subsequent section~\ref{Einstein} we derive the explicit solutions to Einstein's equations in such a setting by performing their step-by-step integration. We summarize the obtained spacetimes and discuss them in section~\ref{sec_discussion}. Appendix~A contains the proof of some useful identities.

\section{General Robinson--Trautman and Kundt geometry}
\label{sec_geom}

In the most natural coordinates the line element of a general non-twisting $D$-dimensional spacetime is given by \cite{PodOrt06}
\begin{equation}
\dd s^2 = g_{pq}(r,u,x)\, \dd x^p\,\dd x^q+2\,g_{up}(r,u,x)\, \dd u\, \dd x^p -2\,\dd u\,\dd r+g_{uu}(r,u,x)\, \dd u^2 \,, \label{general nontwist}
\end{equation}
where $x$ is a shorthand for ${(D-2)}$ spatial coordinates ${x^p}$.\footnote{Throughout this paper the indices $m,n,p,q$ label the spatial coordinates and range from $2$ to ${D-1}$.}
The nonvanishing contravariant metric components are $g^{pq}$ (an inverse matrix to $g_{pq}$), ${g^{ru}=-1}$, ${g^{rp}= g^{pq}g_{uq}}$ and ${g^{rr}= -g_{uu}+g^{pq}g_{up}g_{uq}}$, so that
\begin{equation}
g_{up}= g_{pq}g^{rq} \,, \qquad g_{uu}= -g^{rr}+g_{pq}g^{rp}g^{rq} \,. \label{CovariantMetricComp}
\end{equation}

The geometrically privileged null vector field ${\mathbf{k}=\mathbf{\partial}_r}$ generates a geodesic and affinely parameterized congruence. A direct calculation for the metric (\ref{general nontwist}) immediately shows that the covariant derivative of $\mathbf{k}$ is given by ${k_{a;b}=\Gamma^{u}_{ab}=\frac{1}{2}g_{ab,r}}$, so that ${k_{r;b}=0=k_{a;r}}$. The optical matrix \cite{OrtaggioPravdaPravdova:2013} defined as ${\rho_{ij}\equiv k_{a;b}\,m_i^am_j^b}$ where ${\mathbf{m}_i\equiv m_i^p(g_{up}\partial_r+\partial_p)}$ are ${(D-2)}$ unit vectors forming the orthonormal basis in the transverse Riemannian space is thus simply given by
\begin{equation}
\rho_{ij}=k_{p;q}\,m^p_im^q_j= {\textstyle \frac{1}{2}}g_{pq,r}\,m^p_im^q_j \,. \label{OptMat}
\end{equation}
This can be decomposed into the antisymmetric twist matrix ${A_{ij}\equiv\rho_{[ij]}}$, symmetric traceless shear matrix ${\sigma_{ij}}$ and the trace ${\Theta}$ determining the expansion such that ${\sigma_{ij}+\Theta\delta_{ij}=\rho_{(ij)}}$ with ${\delta^{ij}\sigma_{ij}=0}$, i.e., ${\rho_{ij}=A_{ij}+\sigma_{ij}+\Theta\delta_{ij}}$. From (\ref{OptMat}) we immediately see that ${A_{ij}=0}$ which confirms that the metric (\ref{general nontwist}) is \emph{non-twisting}. If we impose the additional condition that the metric is \emph{shear-free}, ${\sigma_{ij}=0}$,  we obtain the relation ${\Theta\delta_{ij}=\frac{1}{2}g_{pq,r}m^p_im^q_j}$. Using ${g_{pq}m_i^pm_j^q=\delta_{ij}}$ we thus infer
\begin{equation}
g_{pq,r}=2\Theta g_{pq} \,, \quad \hbox{so that} \quad g_{pq,rr}=2\big(\Theta_{,r}+2\Theta^2\big)g_{pq} \,. \label{shearfree condition}
\end{equation}
The first expression can be integrated as
\begin{equation}
{\textstyle g_{pq}=R^2(r,u,x)\,h_{pq}(u,x) \,, \qquad \hbox{where} \qquad R=\exp\big(\int\Theta(r,u,x)\,\dd r\big)} \,. \label{IntShearFreeCond}
\end{equation}
When the expansion vanishes, ${\Theta=0}$, this effectively reduces to ${R=1}$ so that the spatial metric ${g_{pq}(u,x)}$ is independent of the affine parameter $r$. It yields exactly the \emph{Kundt class} of non-expanding, twist-free and shear-free geometries \cite{Stephani:2003,GriffithsPodolsky:2009,PodolskyZofka:2009,OrtaggioPravdaPravdova:2013}. The other case ${\Theta\neq 0}$ gives the expanding \emph{Robinson--Trautman class} which we will study in this contribution.

The Christoffel symbols for the general non-twisting spacetime (\ref{general nontwist}) after applying the shear-free condition (\ref{shearfree condition}) are
\begin{eqnarray}
&& {\textstyle \Gamma^r_{rr} = 0} \,, \label{ChristoffelBegin} \\
&& {\textstyle \Gamma^r_{ru} = -\frac{1}{2}g_{uu,r}+\frac{1}{2}g^{rp}g_{up,r}} \,, \\
&& {\textstyle \Gamma^r_{rp} = -\frac{1}{2}g_{up,r}+\Theta g_{up}} \,, \\
&& {\textstyle \Gamma^r_{uu} = \frac{1}{2}\big[-g^{rr}g_{uu,r}-g_{uu,u}+g^{rp}(2g_{up,u}-g_{uu,p})\big]} \,, \\
&& {\textstyle \Gamma^r_{up} = \frac{1}{2}\big[-g^{rr}g_{up,r}-g_{uu,p}+g^{rq}(2g_{u[q,p]}+g_{qp,u})\big]} \,, \\
&& {\textstyle \Gamma^r_{pq} = -\Theta g^{rr}g_{pq}-g_{u(p||q)}+\frac{1}{2}g_{pq,u}} \,, \\
&& {\textstyle \Gamma^u_{rr}=\Gamma^u_{ru}=\Gamma^u_{rp} = 0} \,, \\
&& {\textstyle \Gamma^u_{uu} = \frac{1}{2}g_{uu,r}} \,, \\
&& {\textstyle \Gamma^u_{up} = \frac{1}{2}g_{up,r}} \,, \\
&& {\textstyle \Gamma^u_{pq} = \Theta g_{pq}} \,, \\
&& {\textstyle \Gamma^m_{rr} = 0} \,, \\
&& {\textstyle \Gamma^m_{ru} = \frac{1}{2}g^{mn}g_{un,r}} \,, \\
&& {\textstyle \Gamma^m_{rp} = \Theta\delta^m_p} \,, \\
&& {\textstyle \Gamma^m_{uu} = \frac{1}{2}\big[-g^{rm}g_{uu,r}+g^{mn}(2g_{un,u}-g_{uu,n})\big]} \,, \\
&& {\textstyle \Gamma^m_{up} = \frac{1}{2}\big[-g^{rm}g_{up,r}+g^{mn}(2g_{u[n,p]}+g_{np,u})\big]} \,, \\
&& {\textstyle \Gamma^m_{pq} = -\Theta g^{rm}g_{pq}+\,^{S}\Gamma^m_{pq}} \,, \label{ChristoffelEnd}
\end{eqnarray}
where ${\,^{S}\Gamma^m_{pq}\equiv\frac{1}{2}g^{mn}(2g_{n(p,q)}-g_{pq,n})}$ are the Christoffel symbols with respect to the spatial coordinates only, i.e., the coefficients of the covariant derivative on the transverse ${(D-2)}$-dimensional Riemannian space.

The Riemann curvature tensor components are then obtained (after straightforward but lengthy calculation) in the form
\begin{eqnarray}
&& R_{rprq} = {\textstyle -\big(\Theta_{,r}+\Theta^2\big)g_{pq}} \,, \\
&& R_{rpru} = {\textstyle -\frac{1}{2}g_{up,rr}+\frac{1}{2}\Theta g_{up,r}} \,, \\
&& R_{ruru} = {\textstyle -\frac{1}{2}g_{uu,rr}+\frac{1}{4}g^{pq}g_{up,r}g_{uq,r}} \,, \\
&& R_{rpmq} = {\textstyle 2g_{p[m}\Theta_{,q]}-2\Theta^2g_{p[m}g_{q]u}+\Theta g_{p[m}g_{q]u,r}} \,, \\
&& R_{rpuq} = {\textstyle \frac{1}{2}g_{up,r||q}+\frac{1}{4}g_{up,r}g_{uq,r}-g_{pq}\Theta_{,u}} \nonumber \\
&& \hspace{15.0mm} {\textstyle -\frac{1}{2}\Theta\big[g_{pq,u}+g_{pq}g_{uu,r}+g_{uq}g_{up,r}-g_{pq}g^{rn}g_{un,r}+2g_{u[p,q]}\big]} \,, \\
&& R_{rupq} = {\textstyle g_{u[p,q],r}+\Theta\big[g_{u[p}g_{q]u,r}-2g_{u[p,q]}\big]} \,, \\
&& R_{ruup} = {\textstyle g_{u[u,p],r}+\frac{1}{4}g^{rn}g_{un,r}g_{up,r}-\frac{1}{2}g^{mn}g_{um,r}E_{np}} \nonumber \\
&& \hspace{15.0mm} {\textstyle +\Theta\big(g_{up,u}-\frac{1}{2}g_{uu,p}-\frac{1}{2}g_{up}g_{uu,r}\big)} \,, \\
&& R_{mpnq} = {\textstyle \,^{S}R_{mpnq}-\Theta^2g^{rr}(g_{mn}g_{pq}-g_{mq}g_{pn})} \nonumber \\
&& \hspace{15.0mm} {\textstyle -\Theta\big(g_{mn}e_{pq}+g_{pq}e_{mn}-g_{mq}e_{pn}-g_{pn}e_{mq}\big)} \,, \\
&& R_{upmq} = {\textstyle g_{p[m,u||q]}+g_{u[q,m]||p}+e_{p[m}g_{q]u,r}} \nonumber \\
&& \hspace{15.0mm} {\textstyle +\Theta\big(g^{rr}g_{p[m}g_{q]u,r}+g_{uu,[q}g_{m]p}-2g^{rn}E_{n[q}g_{m]p}\big)} \,, \\
&& R_{upuq} = {\textstyle -\frac{1}{2}(g_{uu})_{||p||q}+g_{u(p,u||q)}-\frac{1}{2}g_{pq,uu}+\frac{1}{4}g^{rr}g_{up,r}g_{uq,r}} \nonumber \\
&& \hspace{15.0mm} {\textstyle -\frac{1}{2}g_{uu,r}e_{pq}+\frac{1}{2}g_{uu,(p}g_{q)u,r}-g^{rn}E_{n(p}g_{q)u,r}+g^{mn}E_{mp}E_{nq}} \nonumber \\
&& \hspace{15.0mm} {\textstyle -\frac{1}{2}\Theta g_{pq}\big[g^{rr}g_{uu,r}+g_{uu,u}-g^{rn}(2g_{un,u}-g_{uu,n})\big]} \,.
\end{eqnarray}
Finally, the components of the Ricci tensor are
\begin{eqnarray}
&& R_{rr} = {\textstyle -(D-2)\big(\Theta_{,r}+\Theta^2\big)} \,, \label{Ricci rr}\\
&& R_{rp} = {\textstyle -\frac{1}{2}g_{up,rr}-\frac{1}{2}(D-4)\Theta g_{up,r}+g_{up}\Theta_{,r}-(D-3)\Theta_{,p}+(D-2)\Theta^2g_{up}} \,, \label{Ricci rp} \\
&& R_{ru} = {\textstyle -\frac{1}{2}g_{uu,rr}+\frac{1}{2}g^{rp}g_{up,rr}+\frac{1}{2}g^{pq}\big(g_{up,r||q}+g_{up,r}g_{uq,r}\big)} \nonumber \\
&& \hspace{12.0mm} {\textstyle -(D-2)\Theta_{,u}-\frac{1}{2}\Theta\big[g^{pq}g_{pq,u}-(D-4)g^{rp}g_{up,r}+(D-2)g_{uu,r}\big]} \,, \label{Ricci ru} \\
&& R_{pq} = {\textstyle \,^{S}R_{pq}-f_{pq}-g_{pq}\big(g^{rr}\Theta_{,r}-2\Theta_{,u}+2g^{rn}\Theta_{,n}\big)+2g_{u(p}\Theta_{,q)}} \nonumber \\
&& \hspace{12.0mm} {\textstyle +\Theta^2\big[2g_{pq}g^{rn}g_{un}-(D-2)g_{pq}g^{rr}-2g_{up}g_{uq}\big]} \nonumber \\
&& \hspace{12.0mm} {\textstyle +\Theta\big[2g_{u(p||q)}+2g_{u(p}g_{q)u,r}-(D-2)e_{pq}+g_{pq}\big(g_{uu,r}-2g^{rn}g_{un,r}-g^{mn}e_{mn}\big)\big]} \,, \label{Ricci pq} \\
&& R_{up} = {\textstyle -\frac{1}{2}g^{rr}g_{up,rr}-\frac{1}{2}g_{uu,rp}+\frac{1}{2}g_{up,ru}+g^{rn}g_{u[n,p],r}-\frac{1}{2}g^{rn}(g_{up,r||n}+g_{un,r}g_{up,r})} \nonumber \\
&& \hspace{12.0mm} {\textstyle +g^{mn}\big(\frac{1}{2}g_{um,r}g_{un||p}+g_{m[p,u||n]}+g_{u[m,p]||n}-\frac{1}{2}e_{mn}g_{up,r}\big)}\nonumber \\
&& \hspace{12.0mm} {\textstyle +g_{up}\Theta_{,u}+\Theta\big[g_{up}g_{uu,r}+\frac{1}{2}(D-4)(g_{uu}g_{up,r}-g_{uu,p})-g_{up,u}} \nonumber \\
&& \hspace{20.0mm} {\textstyle -g^{rn}g_{un,r}g_{up}+(D-6)g^{rn}(g_{u[n,p]}-\frac{1}{2}g_{un}g_{up,r})+\frac{1}{2}(D-2)g^{rn}g_{np,u}\big]}\,, \label{Ricci up} \\
&& R_{uu} = {\textstyle -\frac{1}{2}g^{rr}g_{uu,rr}-g^{rn}g_{uu,rn}-\frac{1}{2}g^{mn}e_{mn}g_{uu,r}+g^{rn}g_{un,ru}-\frac{1}{2}g^{mn}g_{mn,uu}} \nonumber \\
&& \hspace{12.0mm} {\textstyle +g^{mn}(g_{um,u||n}-\frac{1}{2}g_{uu||m||n})+\frac{1}{2}(g^{rr}g^{mn}-g^{rm}g^{rn})g_{um,r}g_{un,r}} \nonumber \\
&& \hspace{12.0mm} {\textstyle +2g^{mn}g^{rp}g_{um,r}g_{u[n,p]}+\frac{1}{2}g^{mn}g_{um,r}g_{uu,n}+g^{mn}g^{pq}E_{pm}E_{qn}} \nonumber \\
&& \hspace{12.0mm} {\textstyle +\frac{1}{2}\Theta\big[(D-4)g^{rn}(2g_{un,u}-g_{uu,n}-g_{un}g_{uu,r})+(D-2)(g_{uu}g_{uu,r}-g_{uu,u})\big]} \,, \label{Ricci uu}
\end{eqnarray}
and the Ricci scalar is
\begin{eqnarray}
&& R = {\textstyle \,^{S}R+g_{uu,rr}-2g^{rn}g_{un,rr}-2g^{pq}g_{up,r||q}-\frac{3}{2}g^{pq}g_{up,r}g_{uq,r}} \nonumber \\
&& \hspace{8.0mm} {\textstyle +2\Theta_{,r}\big[(D-2)g_{uu}-(D-3)g^{rn}g_{un}\big]+4(D-2)\Theta_{,u}-4(D-3)g^{rn}\Theta_{,n}} \nonumber \\
&& \hspace{8.0mm} {\textstyle -\Theta^2\big[(D-1)(D-2)g^{rr}-2(2D-5)g^{rn}g_{un}\big]} \nonumber \\
&& \hspace{8.0mm} {\textstyle +\Theta\big[2(D-2)g_{uu,r}-2(2D-7)g^{rn}g_{un,r}+(D-1)g^{pq}g_{pq,u}-2(D-3)g^{pq}g_{up||q}\big]} \,.
\end{eqnarray}
In the above expressions, ${\,^{S}R_{mpnq}}$, ${\,^{S}R_{pq}}$ and ${\,^{S}R}$ are the Riemann tensor, Ricci tensor and Ricci scalar for the transverse-space metric ${g_{pq}}$, respectively. The symbol ${\,_{||}}$ denotes the covariant derivative with respect to $g_{pq}$\,:
\begin{eqnarray}
g_{up||q} \erovno g_{up,q}-g_{um}\,^{S}\Gamma^{m}_{pq}  \,, \\
g_{up,r||q} \erovno g_{up,rq}-g_{um,r}\,^{S}\Gamma^{m}_{pq}  \,, \\
g_{p[m,u||q]} \erovno g_{p[m,q],u}+{\textstyle \frac{1}{2}}(\,^{S}\Gamma^{n}_{pm}g_{nq,u}-\,^{S}\Gamma^{n}_{pq}g_{nm,u}) \,,\\
g_{u[q||m]||p} \erovno g_{u[q,m],p}-\,^{S}\Gamma^{n}_{pq}g_{u[n,m]}-\,^{S}\Gamma^{n}_{pm}g_{u[q,n]} \,,\\
(g_{uu})_{||p||q} \erovno g_{uu,pq}-g_{uu,n}\,^{S}\Gamma^{n}_{pq}\,, \\
g_{up,u||q} \erovno g_{up,uq}-g_{um,u}\,^{S}\Gamma^{m}_{pq}  \,,
\end{eqnarray}
and ${e_{pq}}$, ${E_{pq}}$, ${f_{pq}}$ are convenient shorthands defined as
\begin{eqnarray}
e_{pq} \erovno g_{u{(p||q)}}- {\textstyle \frac{1}{2}}g_{pq,u} \,, \\
E_{pq} \erovno g_{u{[p,q]}}+ {\textstyle \frac{1}{2}}g_{pq,u} \,, \\
f_{pq} \erovno g_{u(p,r||q)}+ {\textstyle \frac{1}{2}}g_{up,r}g_{uq,r} \,,
\end{eqnarray}
where, of course, ${g_{u{[p,q]}}=g_{u{[p||q]}}}$. It will also be useful to rewrite the following $r$-derivatives of the metric functions in terms of the contravariant components, see (\ref{CovariantMetricComp}), namely
\begin{eqnarray}
g_{up,r} \rovno g_{pq}\big(g^{rq}_{\hspace{2.4mm},r}+2\Theta g^{rq}\big)\,, \label{gupr} \\
g_{up,rr} \rovno g_{pq}\big(g^{rq}_{\hspace{2.4mm},rr}+2\Theta_{,r} g^{rq}+4\Theta^2g^{rq}+4\Theta g^{rq}_{\hspace{2.4mm},r}\big)\,, \label{guprr} \\
g_{uu,r} \rovno -g^{rr}_{\hspace{2.4mm},r} +2g_{pq}\big(g^{rp}g^{rq}_{\hspace{2.4mm},r}+\Theta g^{rp}g^{rq}\big) \,, \label{guur} \\
g_{uu,rr} \rovno -g^{rr}_{\hspace{2.4mm},rr} +2g_{pq}\big(g^{rp}g^{rq}_{\hspace{2.4mm},rr}+g^{rp}_{\hspace{2.4mm},r}g^{rq}_{\hspace{2.4mm},r}+\Theta_{,r}g^{rp}g^{rq}+2\Theta^2g^{rp}g^{rq}+4\Theta g^{rp}g^{rq}_{\hspace{2.4mm},r}\big) \,.
\end{eqnarray}
The expressions (\ref{Ricci rr})-(\ref{Ricci uu}) of the Ricci tensor enable us to write explicitly the gravitational field equations for any non-twisting and shear-free geometry of an arbitrary dimension $D$, that is for any Kundt or Robinson--Trautman spacetime.

\section{Einstein's field equations with gyratons and their complete integration}
\label{Einstein}

General Einstein's equations for the metric $g_{ab}$ have the form ${R_{ab}-\frac{1}{2}R\,g_{ab}+\Lambda\, g_{ab}=8\pi T_{ab}}$, where we admit a nonvanishing cosmological constant $\Lambda$ and an arbitrary matter field given by its energy momentum-tensor $T_{ab}$ with the trace $T=g^{ab}T_{ab}$. By substituting their trace $R=\frac{2}{D-2}(\Lambda D-8\pi T)$ we obtain
\begin{equation}
R_{ab}= {\textstyle \frac{2}{D-2}\,\Lambda\,g_{ab}+8\pi\big(T_{ab}-\frac{1}{D-2}\,T\,g_{ab}\big)} \,. \label{EinstinEq}
\end{equation}
Our main aim here is to solve the Einstein field equations (\ref{EinstinEq}) in the case of expanding Robinson--Trautman geometry with a \emph{gyratonic matter}, which is a natural generalization of a pure radiation field to admit a spin of the null source \cite{Bonnor:1970b,FrolovFursaev:2005,KrtousPodolskyZelnikovKadlecova:2012}. We thus assume that the only nonvanishing components of the energy-momentum tensor are ${T_{uu}(r,u,x)}$ corresponding to the classical pure radiation component and ${T_{up}(r,u,x)}$ which encodes inner gyratonic angular momentum. We immediately observe from (\ref{general nontwist}), (\ref{CovariantMetricComp}) that the trace of such energy-momentum tensor vanishes, ${T=0}$.

Moreover, the condition ${T^{ab}_{\hspace{2.6mm};b}=0}$ which follows from Bianchi identities, after a straightforward manipulation, gives the constraints
\begin{equation}
T_{up;r}=0 \,, \qquad T_{uu;r}=g^{pq}\,T_{up;q} \,. \\
\end{equation}
These can be explicitly rewritten using (\ref{ChristoffelBegin})--(\ref{ChristoffelEnd}) as
\begin{eqnarray}
&& T_{up,r}-\Theta\,T_{up}=0 \,, \label{EqTup} \\
&& {\textstyle T_{uu,r}+(D-2)\,\Theta\, T_{uu}=g^{pq}\,T_{up||q}+\big[\frac{1}{2}g^{rp}_{\hspace{2.4mm},r}+(D-1)\,\Theta\, g^{rp}\big]T_{up}} \label{EqTuu} \,.
\end{eqnarray}
We can now perform a step-by-step integration of the Einstein field equations (\ref{EinstinEq}).

\subsection{The equation ${R_{rr}= 0}$}
From (\ref{Ricci rr}) we get the explicit form of this equation
\begin{equation}
\Theta_{,r}+\Theta^2=0 \,, \label{feqrr}
\end{equation}
which obviously determines the $r$-dependence of the expansion scalar $\Theta$. Its general solution can be written as $\Theta^{-1}=r+r_{0}(u,x)$. However, the metric (\ref{general nontwist}) is invariant under the gauge transformation ${r \to r-r_{0}(u,x)}$ and we can thus, without loss of generality, set the integration function ${r_{0}(u,x)}$ to zero. The expansion simply becomes
\begin{equation}
\Theta=\frac{1}{r}\,. \label{ExplEx}
\end{equation}
The integral form (\ref{IntShearFreeCond}) of the shear-free condition (\ref{shearfree condition}) with the expansion given by (\ref{ExplEx}) completely determines the $r$-dependence of the ${(D-2)}$-dimensional spatial metric ${g_{pq}(r,u,x)}$, namely
\begin{equation}
g_{pq}=r^2\,h_{pq}(u,x) \,, \label{SpMetr}
\end{equation}
so that ${g^{pq}=r^{-2}h^{pq}}$, where ${h^{pq}}$ is the inverse matrix of ${h_{pq}}$. The $r$-independent metric part ${h_{pq}}$ will be constrained by the next Einstein's equations.

\subsection{The equation ${R_{rp}= 0}$}
Using (\ref{Ricci rp}), (\ref{gupr}) and (\ref{guprr}) we rewrite the Ricci tensor component ${R_{rp}}$ in a more compact way
\begin{equation}
{\textstyle R_{rp}=-\frac{1}{2}g_{pq}\big(g^{rq}_{\hspace{2.4mm},rr}+D\,\Theta\, g^{rq}_{\hspace{2.4mm},r}\big)-(D-3)\,\Theta_{,p}} \,.
\label{feqrp}
\end{equation}
Employing now the restriction given by ${R_{rr}=0}$, i.e., the explicit form of expansion (\ref{ExplEx}) the equation ${R_{rp}= 0}$ becomes
\begin{equation}
g^{rq}_{\hspace{2.4mm},rr}+\frac{D}{r}\,g^{rq}_{\hspace{2.4mm},r}=0 \,.
\end{equation}
We easily find its general solution ${g^{rq}(r,u,x)}$ in the form
\begin{equation}
g^{rq}=e^{q}(u,x)+r^{1-D}f^{q}(u,x) \,, \label{NediagContra}
\end{equation}
where $e^q$ and $f^q$ are arbitrary integration functions of $u$ and $x$. In view of (\ref{CovariantMetricComp}) and (\ref{SpMetr}), the corresponding covariant components of the Robinson--Trautman metric are
\begin{equation}
g_{up}=r^2e_{p}(u,x)+r^{3-D}f_{p}(u,x) \,, \label{NediagCov}
\end{equation}
where ${e_{p}\equiv h_{pq}e^q}$ and ${f_{p}\equiv h_{pq}f^q}$.

At this stage we can also fully integrate the energy-momentum conservation equations (\ref{EqTup}), (\ref{EqTuu}) which determine the $r$-dependence of the gyratonic energy-momentum tensor:
\begin{eqnarray}
&& T_{up} = \mathcal{J}_p\,r \,, \label{ExplTup} \\
&& {\textstyle T_{uu} = \frac{1}{D-2}\,h^{pq}\mathcal{J}_{p||q}+ \mathcal{J}_p\,\big[e^p\,r+\frac{1}{2}(D-1)\,f^p\,r^{2-D}\ln{r}\big]+\mathcal{N}\,r^{2-D}} \,, \label{ExplTuu}
\end{eqnarray}
where ${\mathcal{J}_p(u,x)}$ and ${\mathcal{N}(u,x)}$ are integration functions of $u$ and $x$.

\subsection{The equation ${R_{ru}= -\frac{2}{D-2}\,\Lambda}$}

It is convenient to rewrite the general Ricci tensor component (\ref{Ricci ru}) using the contravariant metric components,
\begin{eqnarray}
&& {\textstyle 2R_{ru}= \,g^{rr}_{\hspace{2.4mm},rr}+(D-2)\,\Theta\,g^{rr}_{\hspace{2.4mm},r}-g_{pq}\big(g^{rp}g^{rq}_{\hspace{2.4mm},rr}+g^{rp}_{\hspace{2.4mm},r}g^{rq}_{\hspace{2.4mm},r}+D\,\Theta\,g^{rp}g^{rq}_{\hspace{2.4mm},r}\big)} \nonumber \\
&&\hspace{12.0mm} {\textstyle +\big(g^{rp}_{\hspace{2.4mm},r}+2\Theta g^{rp}\big)_{||p}-\Theta\,g^{pq}g_{pq,u}-2(D-2)\Theta_{,u}} \,.
\label{feqru}
\end{eqnarray}
Employing the previous results (\ref{ExplEx}), (\ref{SpMetr}) and (\ref{NediagContra}) the corresponding Einstein equation becomes
\begin{eqnarray}
&& {\textstyle g^{rr}_{\hspace{2.4mm},rr}+(D-2)\,r^{-1}\,g^{rr}_{\hspace{2.4mm},r}=-2\big(e^p\,\!_{||p}-\frac{1}{2}h^{pq}h_{pq,u}\big)\,r^{-1}-\frac{4}{D-2}\,\Lambda} \nonumber \\
&&\hspace{42.0mm} {\textstyle +(D-3)f^p\,\!_{||p}\,r^{-D}+(D-1)^2f^pf_p\,r^{2(1-D)}} \,. \label{Rru EEq}
\end{eqnarray}
Its homogeneous solution is ${\,g^{rr}_0=a+b\,r^{3-D}\,}$, where ${a(u,x)}$ and ${b(u,x)}$ are integration functions. The particular solution can be obtained as a superposition of terms ${g^{rr}_{(k)}=\frac{1}{(k+2)(D-1+k)}\,\gamma\,r^{k+2}}$ corresponding to all terms of the form ${\gamma\,r^k}$ on the right hand side of (\ref{Rru EEq}). The general solution with an explicit $r$-dependence of the metric component ${g^{rr}}$ thus becomes
\begin{eqnarray}
&&g^{rr}=a+b\,r^{3-D}-\frac{2}{D-2}\,\big(e^{p}\,\!_{||p}-{\textstyle \frac{1}{2}}h^{pq}h_{pq,u}\big)\,r-\frac{2\Lambda}{(D-1)(D-2)}\,r^2 \nonumber \\
&&\hspace{10.0mm} +\frac{D-3}{D-2}\,f^p\,\!_{||p}\,r^{2-D}+\frac{D-1}{2(D-2)}\,f^pf_p\,r^{2(2-D)} \,. \label{Explicit grr}
\end{eqnarray}
Notice that ${g_{uu}}$ is then simply obtained using (\ref{CovariantMetricComp}) as
\begin{equation}
g_{uu}=-g^{rr}+r^2\,e^pe_p+2\,r^{3-D}\,e^pf_p+r^{2(2-D)}\,f^pf_p \,. \label{guuExpl}
\end{equation}

\subsection{The equation ${R_{pq}=\frac{2}{D-2}\,\Lambda\,g_{pq}}$}

Using (\ref{guur}), (\ref{ExplEx}), (\ref{SpMetr}), (\ref{NediagContra}) and (\ref{NediagCov}) the general Ricci tensor component (\ref{Ricci pq}) becomes
\begin{eqnarray}
&& {\textstyle R_{pq}= \,^{S}R_{pq}-\Big[(D-3)\,g^{rr}+r\,g^{rr}_{\hspace{2.4mm},r}\Big]h_{pq}} \nonumber \\
&&\hspace{10.0mm} {\textstyle -\Big[\big(e^{n}\,\!_{||n}-\frac{1}{2}h^{mn}h_{mn,u}\big)h_{pq}+(D-2)\big(e_{(p||q)}-\frac{1}{2}h_{pq,u}\big)\Big]\,r} \nonumber \\
&&\hspace{10.0mm} {\textstyle +\big(f_{(p||q)}-f^{n}\,\!_{||n}h_{pq}\big)\,r^{2-D}-\frac{1}{2}(D-1)^2f_pf_q\,r^{2(2-D)}} \,,
\end{eqnarray}
in which, employing (\ref{Explicit grr}),
\begin{eqnarray}
&&(D-3)\,g^{rr}+r\,g^{rr}_{\hspace{2.4mm},r}= (D-3)\,a-2\big(e^{n}\,\!_{||n}- {\textstyle \frac{1}{2}}h^{mn}h_{mn,u}\big)\,r-\frac{2\Lambda}{D-2}\,r^2 \nonumber \\
&&\hspace{35.0mm}-\frac{D-3}{D-2}\,f^{n}\,\!_{||n}\,r^{2-D}-\frac{(D-1)^2}{2(D-2)}\,f^nf_n\,r^{2(2-D)} \,. \label{GrrGrrr}
\end{eqnarray}
The corresponding Einstein equations (\ref{EinstinEq}) thus take the form
\begin{eqnarray}
&& \hspace{-5.0mm}\,^{S}R_{pq}-(D-3)\,a\, h_{pq}+\Big[\big(e^{n}\,\!_{||n}- {\textstyle \frac{1}{2}}h^{mn}h_{mn,u}\big)h_{pq}-(D-2)\big(e_{(p||q)}- {\textstyle \frac{1}{2}}h_{pq,u}\big)\Big]\,r \nonumber \\
&&\hspace{5.0mm} + \Big(f_{(p||q)}-\frac{h_{pq}}{D-2}\,f^{n}\,\!_{||n}\Big)\,r^{2-D}-\frac{1}{2}(D-1)^2 \Big(f_pf_q-\frac{h_{pq}}{D-2}\,f^nf_n\Big)\,r^{2(2-D)}=0 \,. \label{EinEq Rpq}
\end{eqnarray}

The trace of this equation explicitly determines the function $a(u,x)$ introduced in (\ref{Explicit grr}), namely
\begin{equation}
a = \frac{\mathcal{R}}{(D-2)(D-3)} \,, \label{aGrr}
\end{equation}
where ${\mathcal{R}\equiv h^{pq}\,\mathcal{R}_{pq}}$ is the Ricci scalar curvature of the spatial metric ${h_{pq}}$ which is the $r$-independent part of ${g_{pq}}$. Notice that due to (\ref{SpMetr}) the corresponding Ricci tensor is ${\mathcal{R}_{pq}\equiv \,^{S}R_{pq}}$, while ${\mathcal{R}\equiv \,^{S}R\,r^2}$. Decomposing the equation (\ref{EinEq Rpq}) into the terms with different powers of $r$ we obtain the following constraints on the metric functions:
\begin{eqnarray}
\mathcal{R}_{pq}&\hspace{-2.0mm}=&\hspace{-2.0mm}\frac{h_{pq}}{D-2}\,\mathcal{R} \,, \label{EinSpace} \\
{\textstyle \frac{1}{2}}h_{pq,u}&\hspace{-2.0mm}=&\hspace{-2.0mm}e_{(p||q)}-\frac{h_{pq}}{D-2}\big(e^{n}\,\!_{||n}- {\textstyle \frac{1}{2}}h^{mn}h_{mn,u}\big) \,, \label{hpquDer} \\
f_{(p||q)}&\hspace{-2.0mm}=&\hspace{-2.0mm}\frac{h_{pq}}{D-2}\,f^{n}\,\!_{||n} \,, \label{fCovDerCond} \\
f_pf_q&\hspace{-2.0mm}=&\hspace{-2.0mm}\frac{h_{pq}}{D-2}\,f^nf_n \,. \label{fCond}
\end{eqnarray}
Now, if we multiply both sides of (\ref{fCond}) by $f^q$ we obtain ${f_p\;(f^qf_q)=\frac{1}{D-2}\,f_p\;(f^qf_q)}$. This necessarily implies
\begin{equation}
f_p=0 \,, \label{fp0}
\end{equation}
whenever ${f^qf_q\neq 0}$. If ${f^qf_q\equiv h^{pq}f_{p}f_{q}=0}$ then again ${f_p=0}$ for all $p$ because the Riemannian metric $h^{pq}$ is a positive definite matrix. In such a case, the condition (\ref{fCovDerCond}) is trivially satisfied.

At this stage, the most general Robinson--Trautman line element (possibly admitting the gyratonic matter) takes the form
\begin{equation}
\dd s^2 = r^2\,h_{pq}\, \dd x^p\dd x^q+2\,r^2\,e_{p}\, \dd u \dd x^p -2\,\dd u\dd r+\big(r^2\,e^pe_p-g^{rr}\big)\, \dd u^2 \,, \label{RTmetric}
\end{equation}
where
\begin{equation}
g^{rr}=a+b\,r^{3-D}+c\,r-{\textstyle\frac{2}{(D-1)(D-2)}}\,\Lambda\,r^2 \,. \label{grr po Rpq}
\end{equation}
with
\begin{equation}
c\equiv -\frac{2}{D-2}\Big(e^{n}\,\!_{||n}-\frac{1}{2}h^{mn}h_{mn,u}\Big)\,. \label{c}
\end{equation}
The functions ${h_{pq}}$ and $e_p$ are constrained by the equations (\ref{EinSpace}) and (\ref{hpquDer}). Due to (\ref{EinSpace}) the transverse ${(D-2)}$-dimensional Riemannian space must be an Einstein space.

\subsection{The equation ${R_{up}=\frac{2}{D-2}\,\Lambda\,g_{up} +8\pi\,T_{up}}$}

Using (\ref{ExplEx}), (\ref{SpMetr}), (\ref{NediagContra}), (\ref{NediagCov}) and (\ref{guuExpl}) with (\ref{fp0}) the Ricci tensor component ${R_{up}}$ (\ref{Ricci up}) becomes
\begin{eqnarray}
&& R_{up}= -e_{p}\Big[(D-3)\,g^{rr}+r\,g^{rr}_{\hspace{2.4mm},r}\Big]+{\textstyle \frac{1}{2}}\Big[g^{rr}_{\hspace{2.4mm},rp}+(D-4)\,r^{-1}g^{rr}_{\hspace{2.4mm},p}\Big] \nonumber \\
&& \hspace{11.0mm}+h^{mn}\big(h_{m[p,u||n]}+e_{[m,p]||n}\big) \nonumber \\
&& \hspace{11.0mm} {\textstyle +\Big[(D-2)\big(e^ne_{[n,p]}-\frac{1}{2}(e^ne_n)_{,p}+\frac{1}{2}e^nh_{np,u}\big)-e_p\big(e^{n}\,\!_{||n}-\frac{1}{2}h^{mn}h_{mn,u}\big)\Big]\,r} \,,
\end{eqnarray}
where
\begin{eqnarray}
&&{\textstyle (D-3)\,g^{rr}+r\,g^{rr}_{\hspace{2.4mm},r}= \frac{1}{D-2}\,\mathcal{R}-2\big(e^{n}\,\!_{||n}-\frac{1}{2}h^{mn}h_{mn,u}\big)\,r-\frac{2}{D-2}\,\Lambda\,r^2} \,, \nonumber \\
&&{\textstyle g^{rr}_{\hspace{2.4mm},rp}+(D-4)\,r^{-1}g^{rr}_{\hspace{2.4mm},p}= -2\,\frac{D-3}{D-2}\big(e^{n}\,\!_{||n}-\frac{1}{2}h^{mn}h_{mn,u}\big)_{,p}} \nonumber \\
&&\hspace{42.0mm} {\textstyle+\frac{(D-4)}{(D-2)(D-3)}\,\mathcal{R}_{,p}\,r^{-1}-b_{,p}\,r^{2-D}} \,,
\end{eqnarray}
see (\ref{Explicit grr}) and (\ref{GrrGrrr}) with (\ref{aGrr}), (\ref{fp0}). The corresponding Einstein equations (\ref{EinstinEq}) with (\ref{ExplTup}) are thus
\begin{eqnarray}
&& {\textstyle -\frac{1}{D-2}\,\mathcal{R}\,e_p-\frac{D-3}{D-2}\big(e^{n}\,\!_{||n}-\frac{1}{2}h^{mn}h_{mn,u}\big)_{,p}+h^{mn}\big(h_{m[p,u||n]}+e_{[m,p]||n}\big)} \nonumber \\
&& {\textstyle +\frac{(D-4)}{2(D-2)(D-3)}\,\mathcal{R}_{,p}\,r^{-1}-\frac{1}{2}\,b_{,p}\,r^{2-D}} \nonumber \\
&& {\textstyle +\Big[(D-2)\big(e^ne_{[n,p]}-\frac{1}{2}(e^ne_n)_{,p}+\frac{1}{2}e^nh_{np,u}\big)+e_p\big(e^{n}\,\!_{||n}-\frac{1}{2}h^{mn}h_{mn,u}\big)\Big]\,r = 8\pi\,\mathcal{J}_p\,r} \,.
\end{eqnarray}
This gives the following conditions:
\begin{eqnarray}
{\textstyle \mathcal{R}\,e_p+(D-3)\big(e^{n}\,\!_{||n}-\frac{1}{2}h^{mn}h_{mn,u}\big)_{,p}-(D-2)h^{mn}\big(h_{m[p,u||n]}+e_{[m,p]||n}\big)=0} \,, && \label{Rup r0} \\
(D-4)\,\mathcal{R}_{,p}=0 \,, && \label{Rup r-1} \\
b_{,p}=0 \,, && \label{Rup r2-D} \\
{\textstyle (D-2)\big(e^ne_{[n,p]}-\frac{1}{2}(e^ne_n)_{,p}+\frac{1}{2}e^nh_{np,u}\big)+e_p\big(e^{n}\,\!_{||n}-\frac{1}{2}h^{mn}h_{mn,u}\big) = 8\pi\,\mathcal{J}_p \label{Rup r} \,.} &&
\end{eqnarray}
Using (\ref{hpquDer}), the relation ${e_{m||p||n}=e_{m||n||p}+e_q\,\mathcal{R}^{q}_{\ mpn}}$ and (\ref{EinSpace}) we find that the equation (\ref{Rup r0}) is satisfied identically. The equation (\ref{Rup r-1}) clearly restricts the dependence of the spatial Ricci scalar ${\mathcal{R}}$ on the spatial coordinates $x^p$, namely
\begin{eqnarray}
&& \mathcal{R}=\mathcal{R}(u)  \qquad \hbox{for} \quad D>4 \,, \label{R D>4}
\label{RDgr4}\\
&& \mathcal{R}=\mathcal{R}(u,x) \quad \hbox{for} \quad D=4 \,. \label{R D=4}
\end{eqnarray}
There is thus a significant difference between the ${D=4}$ case of classical relativity and the extension of Robinson--Trautman spacetimes to higher dimensions. Similarly, the equation (\ref{Rup r2-D}) gives
\begin{equation}
b=b(u) \,. \label{bu}
\end{equation}
Finally, by substituting the expression (\ref{hpquDer}) into the equation (\ref{Rup r}) we get
\begin{equation}
{\textstyle (D-2)\big(e^ne_{n||p}-\frac{1}{2}(e^ne_n)_{,p}\big)=8\pi\,\mathcal{J}_p} \,.
\end{equation}
Since ${(e^ne_n)_{,p}=(e^ne_n)_{||p}}\,$, its left hand side \emph{always vanishes} and we obtain the condition for the energy-momentum tensor (\ref{ExplTup}), (\ref{ExplTuu})
\begin{equation}
\mathcal{J}_p=0 \,.
\label{Jpje0}
\end{equation}
Necessarily, in \emph{any} dimension $D$ we thus obtain
\begin{equation}
T_{up} = 0 \,, \qquad T_{uu} =\mathcal{N}\,r^{2-D} \,, \label{TuuExpl}
\end{equation}
which is just the well-known pure radiation field (null fluid) \emph{without the ``rotational'' components ${T_{up}}$} of the energy-momentum tensor. We have thus proved that there are \emph{no solutions with gyratonic sources in the Robinson--Trautman class of spacetimes}.

\subsection{The equation ${R_{uu}=\frac{2}{D-2}\Lambda\,g_{uu} +8\pi T_{uu}}$}

This final equation determines the relation between the Robinson--Trautman geometry and the pure radiation matter field represented by the profile $\mathcal{N}(u,x)$. Using (\ref{ExplEx}), (\ref{SpMetr}), (\ref{NediagContra}), (\ref{NediagCov}) and (\ref{guuExpl}) with (\ref{fp0}) the Ricci tensor component ${R_{uu}}$ given by (\ref{Ricci uu}) becomes
\begin{eqnarray}
&& R_{uu}= {\textstyle \frac{1}{2}g^{rr}g^{rr}_{\hspace{2.4mm},rr}+\frac{1}{2}\Big[e^{n}\,\!_{||n}-\frac{1}{2}h^{mn}h_{mn,u}+(D-2)\,g^{rr}\,r^{-1}-2\,e^ne_n\,r\Big]g^{rr}_{\hspace{2.4mm},r}} \nonumber\\
&& \hspace{10.0mm} {\textstyle +e^n\Big[g^{rr}_{\hspace{2.4mm},r}+\frac{1}{2}(D-6)\,g^{rr}r^{-1}\Big]_{,n}+\frac{1}{2}h^{mn}g^{rr}_{\hspace{2.4mm}||m||n}\,r^{-2}+\frac{1}{2}(D-2)\,g^{rr}_{\hspace{2.4mm},u}\,r^{-1}} \nonumber \\
&& \hspace{10.0mm} {\textstyle -(D-3)\,e^{n}e_n\,g^{rr}+h^{mn}\Big[e_{m,u||n}-\frac{1}{2}(e^pe_p)_{||m||n}-\frac{1}{2}h_{mn,uu}\Big]} \nonumber \\
&& \hspace{10.0mm} {\textstyle +h^{mn}h^{pq}\big(e_{[p,m]}+\frac{1}{2}h_{pm,u}\big)\big(e_{[q,n]}+\frac{1}{2}h_{qn,u}\big)} \nonumber \\
&& \hspace{10.0mm} {\textstyle +\Big[\frac{1}{2}(D-2)\big(e^me^nh_{mn,u}-e^n(e^pe_p)_{,n}\big)-e^pe_p\big(e^{n}\,\!_{||n}-\frac{1}{2}h^{mn}h_{mn,u}\big)\Big]\,r} \,.
\end{eqnarray}
Moreover, employing the explicit form (\ref{grr po Rpq}) of ${g^{rr}}$ with the help of (\ref{hpquDer}) and (\ref{bu}) we get
\begin{eqnarray}
&& R_{uu}= \textstyle{ \frac{2}{D-2}\,\Lambda g_{uu}+\frac{1}{2}(D-2)\Big[b_{,u}+\frac{1}{2}(D-1)\,b\,c\Big]r^{2-D}+\frac{1}{2}h^{mn}\,a_{||m||n}\,r^{-2}} \nonumber\\
&& \hspace{11.0mm} {\textstyle +\frac{1}{2}\Big[(D-2)\,(a_{,u}+a\,c)+(D-6)\,e^{n}a_{,n}+h^{mn}\,c_{||m||n}\Big]\,r^{-1}} \nonumber\\
&& \hspace{11.0mm} {\textstyle +\frac{1}{2}(D-2)\,(c_{,u}+c^2)+e^{n}\,\!_{||n}\,c+\frac{1}{2}(D-4)\,e^n\,c_{,n}-(D-3)\,e^pe_p\,a} \nonumber\\
&& \hspace{18.0mm} {\textstyle +h^{mn}\Big[e_{m,u||n}-\frac{1}{2}h_{mn,uu}-\frac{1}{2}(e^pe_p)_{||m||n}+h^{pq}e_{p||m}e_{q||n}\Big]} \nonumber\\
&& \hspace{11.0mm} {\textstyle +\frac{1}{2}(D-2)\,\Big[e^me^nh_{mn,u}-e^n(e^pe_p)_{,n}-e^ne_n\,c\Big]\,r} \,,
\label{Ruu1}
\end{eqnarray}
where $a$ is given by (\ref{aGrr}) and $c$ by (\ref{c}). Now, lengthy calculations using the previously derived constraints lead to the following identities (proved in the Appendix):
\begin{eqnarray}
e^me^nh_{mn,u}-e^n(e^pe_p)_{,n}-e^ne_n\,c=0 \,, \hspace{13.4mm}&&  \label{Identity1}\\
{\textstyle \frac{1}{2}(D-2)\,(c_{,u}+c^2)+e^{n}\,\!_{||n}\,c+\frac{1}{2}(D-4)\,e^n\,c_{,n}-(D-3)\,e^pe_p\,a}
\hspace{21.4mm} &&  \nonumber\\
{\textstyle +h^{mn}\big[e_{m,u||n}-\frac{1}{2}h_{mn,uu}-\frac{1}{2}(e^pe_p)_{||m||n}+h^{pq}e_{p||m}e_{q||n}\big]}=0
\,, \hspace{13.4mm}&&  \label{Identity2}\\
{\textstyle (D-2)\,(a_{,u}+a\,c)+(D-6)\,e^{n}a_{,n}+h^{mn}\,c_{||m||n}}=(D-4)\,e^{n}a_{,n} \,, \hspace{-5.5mm}&& \label{Identity3}
\end{eqnarray}
which are exactly the terms in (\ref{Ruu1}) proportional to $r$, $r^0$, and ${r^{-1}}$, respectively. The corresponding Einstein equation ${R_{uu}=\frac{2}{D-2}\Lambda\,g_{uu} +8\pi T_{uu}}$ with (\ref{TuuExpl}) thus takes a very simple form
\begin{equation}
\textstyle{ \frac{1}{2}(D-2)\Big[b_{,u}+\frac{1}{2}(D-1)\,b\,c\Big]r^{2-D}+\frac{1}{2}h^{mn}\,a_{||m||n}\,r^{-2} +\frac{1}{2} (D-4)\,e^{n}a_{,n}\,r^{-1}}
= 8\pi\, \mathcal{N}\,r^{2-D} \,.
\label{RTEq1}
\end{equation}
It is interesting that also the last term on the left-hand side \emph{always vanishes} since ${(D-4)\,a_{,n}=0}$ in \emph{any} dimension $D$, see equation (\ref{Rup r-1}).
Consequently, in any dimension the last Einstein equation can be compactly\footnote{The term proportional to ${r^{-2}}$ in equation (\ref{RTEq1}) is always zero in the case ${D>4}$ since ${a_{||m||n}=0}$ due to (\ref{R D>4}). In the ${D=4}$ case it is combined with the terms proportional to ${r^{2-D}=r^{-2}}$ into the expression (\ref{RTEq}).}
written as
\begin{equation}
\triangle a+{\textstyle\frac{1}{2}}(D-1)(D-2)\,b\,c+(D-2)\,b_{,u}=16\pi\,\mathcal{N} \,, \label{RTEq}
\end{equation}
where ${\triangle a \equiv h^{mn}a_{||m||n}}$ is the covariant Laplace operator on the ${(D-2)}$-dimensional transverse Riemannian space. In particular, this Einstein field equation reads
\begin{eqnarray}
{\textstyle\frac{1}{2}}(D-1)\,b\,c+\,b_{,u} =\frac{16\pi}{D-2}\,\mathcal{N}&&   \quad \hbox{for} \quad D>4 \,, \label{RTEq>4f}\\
\triangle ({\textstyle\frac{1}{2}}\mathcal{R})+3\,b\,c+2\,b_{,u}=16\pi\,\mathcal{N}&&  \quad \hbox{for} \quad D=4 \,. \label{RTEq=4f}
\end{eqnarray}
For the special choice ${e^p=0}$, equation (\ref{RTEq=4f}) reduces exactly to the classical \emph{Robinson--Trautman equation} \cite{Stephani:2003,GriffithsPodolsky:2009} (with the identification ${a={\textstyle\frac{1}{2}}\mathcal{R}=\triangle (\log P)=K}$, ${b=-2m(u)}$,  ${c=-2(\log P)_{,u}}$, where $K$ is the Gaussian curvature of the spatial metric ${h_{pq}=P^{-2}\,\delta_{pq}}$). Equation (\ref{RTEq>4f}) generalizes the field equation previously derived in \cite{PodOrt06} in the sense that now it also includes the contribution from the off-diagonal metric components $e^p$ entering the function $c$ as their  covariant spatial divergence ${e^{n}\,\!_{||n}}$, see (\ref{c}).

\section{Summary and concluding discussion}
\label{sec_discussion}

The most general $D$-dimensional Robinson--Trautman line element in vacuum, with a cosmological constant $\Lambda$ and possibly the pure radiation field ${T_{uu}=\mathcal{N}\,r^{2-D}}$  can thus be written (in the natural gauge in which the expansion is ${\Theta=r^{-1}}$, see (\ref{ExplEx})) as
\begin{equation}
\dd s^2 = r^2\,h_{pq}\, (\dd x^p+e^{p}\, \dd u)(\dd x^q+e^{q}\, \dd u) -2\,\dd u\dd r-g^{rr}\, \dd u^2 \,, \label{RTmetricFin}
\end{equation}
where
\begin{equation}
g^{rr}=\frac{\mathcal{R}}{(D-2)(D-3)}+\frac{b(u)}{r^{D-3}}-\frac{2}{D-2}\,\big(e^{p}\,\!_{||p}-{\textstyle \frac{1}{2}}h^{pq}h_{pq,u}\big)\,r-\frac{2\Lambda}{(D-1)(D-2)}\,r^2 \,,
\label{grrfinal}
\end{equation}
with the functions ${h_{pq}(u,x)}$ and $e^p(u,x)$ constrained by the equations (\ref{EinSpace}), (\ref{hpquDer}), (\ref{RTEq}), namely
\begin{eqnarray}
\mathcal{R}_{pq} = \frac{h_{pq}}{D-2}\,\mathcal{R} \,,&& \label{EinSpacerep} \\
e_{(p||q)}-{\textstyle \frac{1}{2}}h_{pq,u} = \frac{h_{pq}}{D-2}\big(e^{n}\,\!_{||n}- {\textstyle \frac{1}{2}}h^{mn}h_{mn,u}\big)\,,&&  \label{hpquDerrep} \\
\frac{\triangle \mathcal{R}}{(D-2)(D-3)}-(D-1)
\Big(e^{n}\,\!_{||n}-{\textstyle\frac{1}{2}}h^{mn}h_{mn,u}\Big)\,b\,
+(D-2)\,b_{,u}= 16\pi\,\mathcal{N} \,,&& \label{RTEqrep}
\end{eqnarray}
in which $b$ is a function of the null coordinate $u$ only. The first equation (\ref{EinSpacerep}) restricts just the Riemannian metric $h_{pq}$ of the transverse ${(D-2)}$-dimensional space covered by the coordinates~$x^p$, with ${\mathcal{R}_{pq}}$ and ${\mathcal{R}}$ being its Ricci tensor and Ricci scalar. Therefore, \emph{any} Einstein space metric $h_{pq}$ is admitted. The second constraint (\ref{hpquDerrep}) imposes a \emph{specific coupling} between the spatial metric $h_{pq}$ and the off-diagonal metric components represented by ${(D-2)}$ functions ${e^p}$. In addition, there is the Einstein equation (\ref{RTEqrep}) which relates these metric functions and an arbitrary ``mass'' function $b(u)$ to the pure radiation profile $\mathcal{N}$. In the vacuum case ${\mathcal{N}=0}$.

After the step-by-step integration of all Einstein's equations we proved that \emph{ there are no gyratons in the Robinson--Trautman class}. In \emph{any dimension including} ${D=4}$ we necessarily obtained ${\mathcal{J}_p=0}$ so that ${T_{up}=\mathcal{J}_p\,r=0}$, see (\ref{ExplTup}) with  (\ref{Jpje0}) or (\ref{TuuExpl}), which means that the null matter field can not have an ``internal spin'' (angular momentum).

This is in striking contrast to the closely related Kundt family of spacetimes which in general (and in any $D$) admits such gyrating sources, as recently demonstrated in \cite{KrtousPodolskyZelnikovKadlecova:2012} (there is also a comprehensive list of previous works studying particular subclasses of the Kundt gyratons). Such a conclusion is surprising because the Robinson--Trautman family of geometries is the closest to the Kundt family --- both are non-twisting and shear-free, and (at least in ${D=4}$) they admit similar algebraic structures and matter fields.

The questions thus arise about the nature of such a difference and also concerning the possible physical interpretation of the off-diagonal metric functions $e^p$. In the following two short sections we will tackle these two problems.

\subsection{Robinson--Trautman versus gyratonic Kundt spacetimes}
Of course, the Robinson--Trautman class of spacetimes is expanding (${\Theta=r^{-1}\not=0}$) while the Kundt class is non-expanding (${\Theta=0}$). The absence or presence of the gyratons thus must be traced to this geometric difference. In section~\ref{sec_geom} we presented the complete list of all curvature tensor  components  for \emph{any} non-twisting and shear-free geometry which contains both the Robinson--Trautman and the Kundt family. We are thus able to trace the point at which the integration of Einstein's equations with gyratonic energy-momentum tensor starts to differ significantly.

To be specific, by setting ${\Theta=0}$ for the Kundt class in (\ref{shearfree condition}) we immediately obtain ${g_{pq}=h_{pq}}$ independent of $r$, instead of (\ref{SpMetr}) which reads ${g_{pq}=r^2\,h_{pq}}$ in the Robinson--Trautman case. Second field equation (\ref{feqrp}) for ${\Theta=0}$ yields ${g_{up}=e_{p}+r f_{p}}$ instead of (\ref{NediagCov}) which is ${g_{up}=r^2e_{p}+r^{3-D}f_{p}}$. Third field equation (\ref{Ricci ru}) gives, instead of (\ref{Explicit grr}), (\ref{guuExpl}), ${g_{uu}=a+b\,r+[\,\frac{2}{D-2}\,\Lambda+{\textstyle \frac{1}{2}}(f^{p}\,\!_{||p}+f^p f_p)\,]\,r^2}$ in full agreement with equations (64), (65) of \cite{PodolskyZofka:2009}. Apart from different powers of $r$, the metric coefficients for the Kundt and Robinson--Trautman spacetimes thus look very similar.

The main difference between these two types of geometries occurs after employing the next field equation for the spatial Ricci components $R_{pq}$ given by (\ref{Ricci pq}). For the Kundt class this equation is \emph{independent of} $r$, namely
${\mathcal{R}_{pq}= \frac{2}{D-2}\,\Lambda\,h_{pq}+f_{pq}}$,
where ${f_{pq}\equiv f_{(p||q)}+\frac{1}{2}f_pf_q}$. Its trace ${\mathcal{R}= 2\,\Lambda+h^{mn}f_{mn}}$ enables us to rewrite it as
\begin{eqnarray}
\mathcal{R}_{pq}-\frac{h_{pq}}{D-2}\,\mathcal{R}\rovno f_{pq}-\frac{h_{pq}}{D-2}\,h^{mn}f_{mn} \nonumber\\
\rovno \Big(f_{(p||q)}-\frac{h_{pq}}{D-2}\,f^{n}\,\!_{||n}\Big)
+\frac{1}{2}\Big(f_p f_q-\frac{h_{pq}}{D-2}\,f^n f_n\Big) . \label{Kundtpq}
\end{eqnarray}
This imposes a specific coupling between the traceless part of the Ricci curvature $\mathcal{R}_{pq}$ of the $(D-2)$-dimensional Riemannian space and the traceless part of the tensor $f_{pq}$ constructed from the functions $f_p$ determining (part of) the off-diagonal Kundt metric components $g_{up}$\,.

It can now be observed from (\ref{EinEq Rpq}) that exactly the same terms occur in the corresponding field equation for the Robinson--Trautman metric, but with \emph{different powers of} $r$. This is the key point: In the non-expanding Kundt class we have obtained just one condition (\ref{Kundtpq}) whereas in the expanding Robinson--Trautman class there are \emph{four separate constraints}, namely (\ref{EinSpace})--(\ref{fCond}). It is the severe constraint (\ref{fCond}) that necessarily requires ${f_p=0}$, see (\ref{fp0}), which then fulfills (\ref{fCovDerCond}) identically. In the Robinson--Trautman case we are thus left with the condition (\ref{EinSpace}) which for ${f_p=0}$ is the \emph{same as} (\ref{Kundtpq}) in the non-expanding case. However, in the Robinson--Trautman case there is the \emph{additional constraint} (\ref{hpquDer}), i.e., (\ref{hpquDerrep}) that couples ${e_{(p||q)}-\frac{1}{2}h_{pq,u}}$ to its trace. It turns out that this specific restriction on possible functions $e_p$ determining the other part of the off-diagonal metric components $g_{up}$ forbids --- after applying the following Einstein's equation for $R_{up}$ --- the presence of gyratonic matter fields in the Robinson--Trautman geometries, see (\ref{Jpje0}). In contrast, there is \emph{no such constraint on} $e_p$ in the non-expanding family which enables gyratons to be included in the Kundt geometries.

\subsection{Robinson--Trautman gravitational waves in ${D=4}$}
Finally, we will elucidate the physical meaning of the functions $e_p$. Instead of representing angular momentum of a gyratonic matter they directly encode amplitudes of the Robinson--Trautman gravitational waves. In usual ${D=4}$ dimensions, the transverse Riemannian space is 2-dimensional. If such a 2-space has \emph{constant curvature}, its metric $h_{pq}$  can be written in the conformally flat form
\begin{equation}
\dd s_0^2 = h_{pq}\, \dd x^p\,\dd x^q = \psi^{-2}\left[(\dd x^2)^2+(\dd x^3)^2\right]\,,\quad\hbox{where}\quad
\psi\equiv 1+{\textstyle\frac{1}{2}}\,\epsilon\,[(x^2)^2+(x^3)^2]\,, \label{spatial metric}
\end{equation}
with ${\epsilon=0, +1 \hbox{ or } -1}$. For such a metric the Christoffel symbols are
\begin{equation}
\qquad
\,^{S}\Gamma^2_{22}=\,^{S}\Gamma^3_{23}=-\,^{S}\Gamma^2_{33}=-\epsilon \, x^2\, \psi^{-1}  \,,\qquad
\,^{S}\Gamma^3_{33}=\,^{S}\Gamma^2_{32}=-\,^{S}\Gamma^3_{22}=-\epsilon \, x^3\, \psi^{-1}  \,. \label{2Da}
\end{equation}
Non-trivial Riemann and Ricci tensor components read ${\mathcal{R}_{2323}=2\,\epsilon\,\psi^{-4}}$, ${\mathcal{R}_{22}=\mathcal{R}_{33}=2\,\epsilon\,\psi^{-2}}$,  so that the Ricci scalar is ${\mathcal{R}=4\,\epsilon=2\,K}$, where $K$ is the (constant) Gaussian curvature. This obviously satisfies the constraint (\ref{EinSpacerep}). It remains to fulfill the constraint (\ref{hpquDerrep}). Since $h_{pq}$  given by (\ref{spatial metric}) is independent of $u$, it reduces to ${e_{(p||q)}=\frac{1}{2}h_{pq}\,e^{n}\,\!_{||n}}$ which is, using (\ref{2Da}),
\begin{eqnarray}
e_{(2||2)} \rovno \psi^{-2}\,{e^2}_{,2}-\epsilon\, \psi^{-3}\,(x^2e^2+x^3e^3)\,, \nonumber\\
e_{(3||3)} \rovno \psi^{-2}\,{e^3}_{,3}-\epsilon\, \psi^{-3}\,(x^2e^2+x^3e^3)\,,\nonumber\\
e_{(2||3)} \rovno {\textstyle\frac{1}{2}}\psi^{-2}\,({e^2}_{,3}+{e^3}_{,2})\,,\nonumber\\
e^{n}\,\!_{||n} \rovno ({e^2}_{,2}+{e^3}_{,3}) -2\,\epsilon\,\psi^{-1}\,(x^2e^2+x^3e^3)\,.
\label{2constr}
\end{eqnarray}
The constraint is thus equivalent to very simple two conditions\footnote{They imply that $e^2$ and $e^3$ are harmonic conjugate functions in \emph{flat} 2-space, ${\triangle e^2=0=\triangle e^3}.$}
\begin{equation}
{e^2}_{,2}={e^3}_{,3} \qquad \hbox{and} \qquad {e^2}_{,3}=-{e^3}_{,2}\,. \label{CauchyRiemann}
\end{equation}
Clearly, these are just the Cauchy--Riemann conditions for the \emph{complex} function $f$ constructed from the real functions $e^2(u,x^2,x^3)$ and $e^3(u,x^2,x^3)$, which depend on an external parameter $u$ and the complex variable $\xi$ composed from the spatial coordinates $x^2$ and $x^3$. In particular, introducing the complex quantities
\begin{equation}
\xi\equiv{\textstyle\frac{1}{\sqrt2}}(x^2+ \hbox{i}\, x^3)\,, \qquad \hbox{and} \qquad
 f \equiv -{\textstyle\frac{1}{\sqrt2}}(e^2+ \hbox{i}\, e^3)\,, \label{complex}
\end{equation}
we obtain that $f$ is a \emph{holomorphic function} of the complex variable $\xi$ since (\ref{CauchyRiemann}) is equivalent to ${f_{,\bar\xi}=0}$ while ${f_{,\xi}=-({e^2}_{,2}+ \hbox{i}\, {e^3}_{,2})}$. \emph{Any} complex function $f(u, \xi)$ holomorphic in $\xi$ thus automatically satisfies the constraint (\ref{hpquDerrep}).

Therefore, in Einstein's ${D=4}$ general relativity it is convenient to adopt the complex representation $\xi$ of the spatial coordinates in the transverse 2-space and the complex function $f(\xi)$ to represent the off-diagonal metric functions $e^2$ and $e^3$. Performing the transformation (\ref{complex}), the general vacuum Robinson--Trautman solution (\ref{RTmetricFin}), (\ref{grrfinal}) (possibly with a cosmological constant $\Lambda$ and/or pure radiation field) whose transverse 2-space has constant curvature  takes the form
\begin{equation}
\dd s^2 = 2\,\frac{r^2}{\psi^2}\, |\dd \xi-f(u, \xi) \, \dd u|^2 -2\,\dd u\dd r-g^{rr}\, \dd u^2 \,, \label{RTmetricFinComplex}
\end{equation}
${\psi\equiv 1+\epsilon\,\xi\bar{\xi}}$ with ${\epsilon=0}$, $+1$, or $-1$, and
\begin{equation}
g^{rr}=2\,\epsilon+\frac{b}{r}
+\Big((f_{,\xi}+\bar{f}_{,\bar{\xi}})-\frac{2\,\epsilon}{\psi}(\bar{\xi} f +\xi\bar{f})\Big)\,r-\frac{\Lambda}{3}\,r^2 \,.
\label{grrfinalComplex}
\end{equation}
Here ${b=b(u)}$ is an arbitrary function and we used the fact that
\begin{equation}
e^{p}\,\!_{||p}=
-(f_{,\xi}+\bar{f}_{,\bar{\xi}})+\frac{2\,\epsilon}{\psi}(\bar{\xi} f +\xi\bar{f})\,.
\label{dive}
\end{equation}
The field equation (\ref{RTEqrep}) is now reduced to a simple relation
${2\,b_{,u}-3\,b\,e^{p}\,\!_{||p}=16\pi\,\mathcal{N}}$
which in the \emph{vacuum case} is just
\begin{equation}
2\,b_{,u}=3\,b\,e^{p}\,\!_{||p}\,. \label{RTEq=4fredvac}
\end{equation}

For ${b=0}$ this vacuum field equation is identically satisfied. In such a case we obtain the complete family of \emph{Robinson--Trautman gravitational waves of algebraic type}~N. Indeed, the metric (\ref{RTmetricFinComplex}), (\ref{grrfinalComplex}) with ${b=0}$ is exactly the line element written in section 4 of \cite{PodolskyGriffiths:2004} above equation (16).
This is related to the Garc\'{\i}a D\'{\i}az--Pleba\'nski 1981 form \cite{GarciaPlebanski:1981} of these exact radiative spacetimes,
\begin{equation}
\dd s^2=2\,v^2\dd\xi\dd\bar\xi+2\,v\bar {A}\,\dd\xi\dd u+2\,v A\,\dd\bar\xi \dd u
+ 2\, \psi\, \dd u\dd v+2\,(A\bar {A}+\psi B)\,\dd u^2\,,
\label{RTmetrGDP}
\end{equation}
where ${A\equiv\epsilon\xi-v f}$,
${B\equiv-\epsilon+\frac{1}{2}v\,(f_{,\xi}+\bar f_{,\bar\xi})+\frac{1}{6}\Lambda v^2\psi}$,
via a simple transformation ${r=-v\psi}$, see also \cite{BicakPodolsky:1999a}.
The corresponding gravitational wave amplitudes ${\cal A}_+$ and ${\cal A}_\times$ of the two independent polarizations are directly determined by the \emph{second covariant derivatives of the function} ${c=-e^{p}\,\!_{||p}}$, cf. (\ref{c}), namely by the symmetric traceless ${2\times2}$ matrix
\begin{equation}
w_{pq}\equiv c_{||p||q}-{\textstyle\frac{1}{2}}h_{pq}\,h^{mn}c_{||m||n}\,.
\label{Wpq}
\end{equation}
Indeed, using (\ref{2Da}) it follows that
\begin{eqnarray}
w_{22} \rovno -w_{33}={\textstyle\frac{1}{2}}(c_{,2,2}-c_{,3,3})+\epsilon\, \psi^{-1}\,(x^2c_{,2}-x^3c_{,3})\,, \nonumber\\
w_{23} \rovno w_{32}=c_{,2,3}+\epsilon\, \psi^{-1}\,(x^3c_{,2}+x^2c_{,3})\,,
\label{Wpqreal}
\end{eqnarray}
which can be rewritten in the complex notation (\ref{complex}) as
${w_{33}=-2\, {\cal R}e \Psi}$,  ${w_{23}=-2\, {\cal I}m \Psi}$, where
${\Psi\equiv{\textstyle\frac{1}{2}\psi^{-2}}\,(\psi^2 c_{,\xi})_{,\xi}}$.
Substituting for ${c=-e^{p}\,\!_{||p}}$ from (\ref{dive}) we immediately obtain ${\Psi=\frac{1}{2}\,f_{,\xi\xi\xi}}$. This yields very simple explicit relations
\begin{equation}
w_{33}=-{\cal R}e \,f_{,\xi\xi\xi}\,,\quad w_{23}=-{\cal I}m \,\,f_{,\xi\xi\xi}\,.
\label{Wpqcomplex}
\end{equation}
By comparing with the expressions (34) of \cite{BicakPodolsky:1999b} determining the two amplitudes of the Robinson--Trautman gravitational waves (measured by a geodesic deviation in a suitable orthonormal frame) we observe that ${{\cal A}_+\propto w_{33}}$ and ${{\cal A}_\times \propto w_{23}}$.

We thus conclude that although the off-diagonal metric components ${g_{up}=r^2e_p}$ can  be locally removed from the metric (\ref{RTmetricFin}) by a gauge transformation ${x'(x,u)}$ such that ${\dd x'^p=\dd x^p+e^p\,\dd u}$, it is in fact \emph{convenient} to keep $e^p$ (or, equivalently, the complex function $f$) non-trivial because these functions directly encode the amplitudes of the Robinson--Trautman gravitational waves. The physical meaning of these metric components in higher dimensions remains an open question since various independent results indicate that there are no Robinson--Trautman gravitational waves in ${D>4}$ \cite{PodOrt06,OrtaggioPravdaPravdova:2013,PravdaPravdovaColeyMilson:2004,SvarcPodolsky:2014b}.

\section*{Acknowledgments}
J.P. has been supported by the grant GA\v{C}R P203/12/0118 and R.\v{S}. by the Czech--Austrian MOBILITY grant
7AMB13AT003.

\newpage
\section*{Appendix~A. Proof of the useful identities}
\renewcommand{\theequation}{A\arabic{equation}}
\setcounter{equation}{0}

Here we present the steps which enable us to prove the non-trivial identities (\ref{Identity1})--(\ref{Identity3}).

\subsection*{Identity (\ref{Identity1})}
This identity immediately follows from the constraint (\ref{hpquDer}) which, in view od the definition (\ref{c}), can be written as
\begin{equation}
h_{mn,u}=2\,e_{(m||n)}+h_{mn}\,c \,. \label{hpquDerc}
\end{equation}
Multiplying this equation by ${e^me^n}$ we obtain ${e^me^nh_{mn,u}=2\,e^ne^me_{m||n}+e^ne_n\,c}$ which is equal to
\begin{equation}
e^me^nh_{mn,u}=e^n(e^me_m)_{,n}+e^ne_n\,c \,. \label{hpquDercid}
\end{equation}

\subsection*{Identity (\ref{Identity2})}
First, it can be shown using (\ref{c}), (\ref{hpquDerc}) and the relation ${{h^{mn}}_{,u}=-h^{mp}h^{nq}h_{pq,u}}$ that
\begin{equation}
{\textstyle \frac{1}{2}(D-2)\,(c_{,u}+c^2)+e^{n}\,\!_{||n}\,c
+h^{mn}\big[e_{m,u||n}-\frac{1}{2}h_{mn,uu}\big]=
h^{mn}\big[e_{m,u||n}-e_{m||n,u}\big]}\,. \label{Identity2a}
\end{equation}
Moreover, using the explicit expressions for the spatial covariant derivatives
\begin{eqnarray}
e_{m,u||n} \rovno e_{m,un}-e_{p,u}\,^{S}\Gamma^{p}_{mn}  \,, \\
e_{m||n,u} \rovno \left(e_{m,n}-e_{p}\,^{S}\Gamma^{p}_{mn}\right)_{,u}  \,, \\
e_{(p||m)||n} \rovno e_{(p||m),n}-e_{(q||m)}\,^{S}\Gamma^{q}_{pn}-e_{(p||q)}\,^{S}\Gamma^{q}_{mn} \,, \label{expr3}
\end{eqnarray}
we obtain
\begin{equation}
{\textstyle
h^{mn}\big[e_{m,u||n}-e_{m||n,u}\big]=-\frac{1}{2}(D-4)\,e^n\,c_{,n}
+e^ph^{mn}e_{p||m||n}+e^ph^{mn}\big[e_{m||p||n}-e_{m||n||p}\big]
} \,. \label{Identity2b}
\end{equation}
Simple calculation yields
\begin{equation}
{\textstyle
h^{mn}\big[-\frac{1}{2}(e^pe_p)_{||m||n}+h^{pq}e_{p||m}e_{q||n}\big]=
-e^ph^{mn}e_{p||m||n}} \,. \label{Identity2c}
\end{equation}
Putting (\ref{Identity2a}), (\ref{Identity2b}) and (\ref{Identity2c}) together, applying the  definition ${e_{m||p||n}-e_{m||n||p}\equiv -{\mathcal{R}^q}_{mnp}\,e_q}$ and using the constraint
${\mathcal{R}_{pq}=(D-3)\,h_{pq}\,a}$, which follow from (\ref{aGrr}), (\ref{EinSpace}),
we thus prove
\begin{eqnarray}
{\textstyle \frac{1}{2}(D-2)\,(c_{,u}+c^2)+e^{n}\,\!_{||n}\,c+\frac{1}{2}(D-4)\,e^n\,c_{,n}-(D-3)\,e^pe_p\,a}
\hspace{8.0mm} &&  \nonumber\\
{\textstyle +h^{mn}\big[e_{m,u||n}-\frac{1}{2}h_{mn,uu}-\frac{1}{2}(e^pe_p)_{||m||n}+h^{pq}e_{p||m}e_{q||n}\big]}=0
\,. &&  \label{Identity2d}
\end{eqnarray}
which is the identity (\ref{Identity2}).

\subsection*{Identity (\ref{Identity3})}
Using the explicit form (\ref{aGrr}) of $a$, where ${\mathcal{R}=h^{pq}\,\mathcal{R}_{pq}}$, (\ref{EinSpace}) and (\ref{c}) it can be shown that
\begin{equation}
{\textstyle (D-2)\,(a_{,u}+a\,c)=\frac{1}{D-3}\,h^{pq}\,\mathcal{R}_{pq,u}-2\,a\,e^{n}\,\!_{||n}} \,. \label{aDiffu}
\end{equation}
It remains to evaluate the term ${h^{pq}\,\mathcal{R}_{pq,u}}$ which (from the definition of the Ricci tensor) is
\begin{equation}
{\textstyle h^{pq}\,\mathcal{R}_{pq,u}=h^{pq}\big[\,^{S}\Gamma^m_{pq,mu}-\,^{S}\Gamma^m_{pm,qu}+\,^{S}\Gamma^m_{pq,u}\,^{S}\Gamma^n_{mn}
  +\,^{S}\Gamma^n_{pq}\,^{S}\Gamma^m_{nm,u}-2\,^{S}\Gamma^m_{pn,u}\,^{S}\Gamma^n_{mq}\big]}\,.
\label{defRicciu}
\end{equation}
Direct calculation using the identity ${{h^{mn}}_{,u}=-h^{mp}h^{nq}h_{pq,u}}$ followed by relations (\ref{hpquDerc}), (\ref{expr3}) and ${e_{m||p||n}-e_{m||n||p}= -{\mathcal{R}^q}_{mnp}\,e_q}$ reveals that
\begin{equation}
T^m_{pq}\equiv{\textstyle \,^{S}\Gamma^m_{pq,u}=h^{mn}\big[e_{n||(p||q)}-e^k\,\mathcal{R}_{k(pq)n}\big]+\delta^m_{\ (p}\,c_{,q)}-\frac{1}{2}h^{mn}h_{pq}\,c_{,n}} \,.
\label{Ttens}
\end{equation}
It is important to observe that the quantity $T^m_{pq}$ is a \emph{tensor in the transverse $(D-2)$-dim space}. Therefore, re-expressing (\ref{defRicciu}) using the covariant derivative $T^m_{pq||m}$ we get the tensor relation
\begin{equation}
{\textstyle h^{pq}\,\mathcal{R}_{pq,u}=h^{pq}\big[\,T^m_{pq||m}-T^m_{pm||q}\,\big]}\,.
\label{defRicciutens}
\end{equation}
Substituting (\ref{Ttens}) into (\ref{defRicciutens}) we obtain
\begin{equation}
h^{pq}\,\mathcal{R}_{pq,u}=
h^{mn}h^{pq}\big[e_{n||p||q||m}-e_{n||p||m||q}\big]+
2h^{mn}\big(e^{p}\,\mathcal{R}_{pm}\big)_{||n}-(D-3)h^{mn}\,c_{||m||n} \,. \label{RpqDiffuexpl}
\end{equation}
Now, the contraction of the identity (3.2.21) of \cite{Wald:1984} yields the identity
\begin{equation}
h^{mn}h^{pq}\big[e_{n||p||q||m}-e_{n||p||m||q}\big]=0\,,
\label{Identity4}
\end{equation}
while the direct evaluation using (\ref{aGrr}), (\ref{EinSpace}) gives
\begin{equation}
2h^{mn}\big(e^{p}\,\mathcal{R}_{pm}\big)_{||n}=2(D-3)\big[e^n\,a_{,n}+a\,e^{n}\,\!_{||n}\big] \,.
\label{Identity5}
\end{equation}
Putting (\ref{RpqDiffuexpl}) with (\ref{Identity4}), (\ref{Identity5}) into  (\ref{aDiffu}) we finally obtain the identity
\begin{equation}
{\textstyle (D-2)\,(a_{,u}+a\,c)+(D-6)\,e^{n}a_{,n}+h^{mn}\,c_{||m||n}}=(D-4)\,e^{n}a_{,n} \,,
\end{equation}
which completes the proof.


\begin{thebibliography}{10}


\bibitem{RobTra60}
I.~Robinson and A.~Trautman, Spherical gravitational waves,
Phys. Rev. Lett. {\bf 4} (1960) 431--432.

\bibitem{RobTra62}
I.~Robinson and A.~Trautman, Some spherical gravitational waves in general relativity,
Proc. Roy. Soc.~A {\bf 265} (1962) 463--473.

\bibitem{Stephani:2003}
H.~Stephani, D.~Kramer, M.~MacCallum, C.~Hoenselaers, and E.~Herlt, {\em Exact Solutions of Einstein's Field Equations}
(Cambridge University Press, Cambridge, 2003).

\bibitem{GriffithsPodolsky:2009}
J.~B.~Griffiths and J.~Podolsk\'{y}, {\em Exact Space-Times in Einstein's General Relativity}
(Cambridge University Press, Cambridge, 2009).

\bibitem{PodOrt06}
J.~Podolsk\'y and M.~Ortaggio, Robinson--Trautman spacetimes in higher dimensions,
Class. Quantum Grav. {\bf 23} (2006) 5785--5797.

\bibitem{Ortaggio07}
M.~Ortaggio, Higher dimensional spacetimes with a geodesic, shearfree, twistfree and expanding null congruence,
Proc. 17th SIGRAV Conf. arXiv:gr-qc/0701036.

\bibitem{OrtPodZof08}
M.~Ortaggio, J.~Podolsk\'y and M.~\v{Z}ofka, Robinson--Trautman spacetimes with an electromagnetic field in higher dimensions,
Class. Quantum Grav. {\bf 25} (2008) 025006 (18pp).

\bibitem{OrtaggioPravdaPravdova:2013}
M.~Ortaggio, V.~Pravda and A.~Pravdov\'a, Algebraic classification of higher dimensional spacetimes based on null alignment,
Class. Quantum Grav. {\bf 30} (2013) 013001 (57pp).

\bibitem{PodolskyZofka:2009}
J.~Podolsk\'{y} and M.~\v{Z}ofka, General Kundt spacetimes in higher dimensions,
Class. Quant. Grav. {\bf 26} (2009) 105008 (18pp).

\bibitem{Bonnor:1970b}
W.~B.~Bonnor, Spinning null fluid in general relativity,
Int. J.~Theor. Phys. {\bf 3} (1970) 257--266.

\bibitem{Griffiths:1972}
J.~B.~Griffiths, Some physical properties of neutrino-gravitational fields,
Int. J.~Theor. Phys. {\bf 5} (1972) 141--150.

\bibitem{FrolovFursaev:2005}
V.~P.~Frolov and D.~V.~Fursaev, Gravitational field of a spinning radiation beam pulse in higher dimensions,
Phys. Rev.~D {\bf 71} (2005) 104034 (16pp).

\bibitem{FrolovIsraelZelnikov:2005}
V.~P.~Frolov, W.~Israel, and A.~Zelnikov, Gravitational field of relativistic gyratons,
Phys. Rev.~D {\bf 72} (2005) 084031 (11pp).

\bibitem{FrolovZelnikov:2005}
V.~P.~Frolov and A.~Zelnikov, Relativistic gyratons in asymptotically AdS spacetime,
Phys. Rev.~D {\bf 72} (2005) 104005 (10pp).

\bibitem{FrolovZelnikov:2006}
V.~P.~Frolov and A.~Zelnikov, Gravitational field of charged gyratons,
Class. Quantum Grav. {\bf 23} (2006) 2119--2128.

\bibitem{KrtousPodolskyZelnikovKadlecova:2012}
P.~Krtou\v{s}, J.~Podolsk\'{y}, A.~Zelnikov and H.~Kadlecov\'{a}, Higher-dimensional Kundt waves and gyratons,
Phys. Rev.~D {\bf 86} (2012) 044039 (17pp).

\bibitem{Wald:1984}
R.~M.~Wald, {\em General Relativity}
(University of Chicago Press, Chicago, 1984).

\bibitem{PodolskyGriffiths:2004}
J.~Podolsk\'y and J.~B.~Griffiths, A snapping cosmic string in a de Sitter or anti-de Sitter universe,
Class. Quantum Grav. {\bf 21} (2004) 2537--2547.

\bibitem{GarciaPlebanski:1981}
A.~Garc\'{\i}a D\'{\i}az and J.~F.~Pleba\'nski, All non-twisting N's with cosmological constant,
J.~Math. Phys. {\bf 22} (1981) 2655--2658.

\bibitem{BicakPodolsky:1999a}
J.~Bi\v{c}\'ak and J.~Podolsk\'y, Gravitational
waves in vacuum spacetimes with cosmological constant. I. Classification and
geometrical properties of non-twisting type N solutions,
J.~Math. Phys. {\bf 40} (1999) 4495--4505.

\bibitem{BicakPodolsky:1999b}
J.~Bi\v{c}\'ak and J.~Podolsk\'y, Gravitational
waves in vacuum spacetimes with cosmological constant. II. Deviation of
geodesics and interpretation of non-twisting type N solutions,
J.~Math. Phys. {\bf 40} (1999) 4506--4517.

\bibitem{PravdaPravdovaColeyMilson:2004}
V.~Pravda, A.~Pravdov\'a, A.~Coley and R.~Milson, Bianchi identities in higher dimensions,
Class. Quantum Grav. {\bf 21} (2004) 2873--2897.

\bibitem{SvarcPodolsky:2014b}
J.~Podolsk\'{y} and R.~\v{S}varc, Algebraic structure of Robinson--Trautman and Kundt geometries in arbitrary dimension, in preparation.

\end{thebibliography}
\end{document}